# Low-Complexity Channel Reconstruction Methods Based on SVD-ZF Precoding in Massive 3D-MIMO Systems

Yuwei Ren[1], Yang Song[2], Xin Su[2]
[1]Beijing University of Posts and Telecommunications, Beijing, China
[2]State Key Laboratory of Wireless Mobile Communications, China Academy of Telecommunications Technology, Beijing, China

**Abstract**: In this paper, we study the low-complexity channel reconstruction methods for downlink precoding in massive multiple-Input multiple-Output (MIMO) systems. When the user is allocated less streams than the number of its antennas, the base station (BS) or user usually utilizes the singular value decomposition (SVD) to get the effective channels, whose dimension is equal to the number of streams. This process is called channel reconstruction and done in BS for time division duplex (TDD) mode. However, with the increasing of antennas in BS, the computation burden of SVD is getting incredible. Here, we propose a series of novel low-complexity channel reconstruction methods for downlink precoding in 3D spatial channel model. We consider different correlations between elevation and azimuth antennas, and divide the large dimensional matrix SVD into two kinds of small-size matrix SVD. The simulation results show that the proposed methods only produce less than 10% float computation than the traditional SVD zero-forcing (SVD-ZF) precoding method, while keeping nearly the same performance of 1Gbps.
**Key words**: Channel Reconstruction; SVD; 3D-MIMO; Massive MIMO;

## I. INTRODUCTION

Massive MIMO is one of the most importantly investigated subjects in the literature of coming fifth-generation (5G) technologies due to the high potential it offers in improving not only the reliability but also the throughput of the system. In fact, Information theory has shown that the capacity of Multi-User MIMO (MU-MIMO) channels could be achieved through simple zero-forcing (ZF) methods [1]. Meanwhile, plane antennas and 3D spatial channel model make massive MIMO implemented possible.

In the past, the traditional arrays only consider the azimuth domain, widely called uniform linear array (ULA). Recently, uniform plane array (UPA) is proposed to cope with 3D channel model. And significant interest has risen on exploring the new spatial character of the elevation / azimuth domain in UPA, in order to devise efficient MIMO transmission / reception schemes. In UPA, passive or active antennas are placed in the elevation domain to provide additional degrees of freedom, which are effectively generated by elevation spread of the channels and the elevation angle distribution of users.

The elevation domain antennas are used to form multiple elevation domain beams, just as Fig.1 shown. For example, these beams can be used to form subsectors in the elevation domain across the system bandwidth, where the subsectors adapt slowly depending on the user load and locations. In another example, these elevation domain beams are frequency-selective and can adapt quickly to cope with various user traffic and conditions, which allows fully flexible multi-user MIMO exploiting the full three dimensional (3D) channel. Some recent deployments taking advantage of elevation domain beams (e.g., [2, 3]), have shown up to 30 percent gain in system capacity.

Despite such huge performance potential, the deployment in realistic setup is hindered by several practical challenges that are not of concern in conventional MIMO systems. The computational complexity is a main concern when the numbers of antennas and users go to be large or infinite. For example, the computation burden of matrix decomposition in channel reconstruction would be a big challenge. The traditional singular value decomposition (SVD) usually requires $O(Nt^2)$ float computations [4], where $Nt$ is the number of BS antennas. With $Nt$ increasing, this is an inconceivable challenge for hardware implementation.

Some work has been completed in the face of such challenge. For example, [5] [6] give some iterative convergence algorithms to approximate matrix SVD, widely used in traditional MIMO systems. [5] adopts randomized algorithms to construct an approximate SVD only with less than 5 iterations, especially for approximating an input matrix with a low-rank element. But this method does not cut the affect of $Nt$, and with the BS antennas increasing, the convergence cannot be proved in limited iteration, which might lead to larger matrix computation than traditional SVD [4]. Besides, 3GPP is actively developing the 3D channel model to enable the evaluation of elevation beamforming. For this, [7][8] apply the 3D beamforming to massive MIMO, where the elevation and azimuth antennas make beamforming respectively, and the two beamforming vectors are transformed into precoding matrix by Kronecker product. Simply, the two beamforming vectors could be the main singular-vectors of different two dimensions, and the corresponding precoding matrix is maximum ratio transmission (MRT). Note that my paper takes similar idea partly.

This paper proposes a series of channel reconstruction methods based on the special structure of 3D spatial channel model. The novelty mainly includes three parts:

➢ Firstly, our methods are fully to explore the special characters of 3D-MIMO system. They consider the different correlations between elevation and azimuth antennas, as well as the different angle resolution between elevation and azimuth directions, and divide the large dimensional SVD into two kinds of small-size SVD.

➢ Secondly, our methods cut the matrix iterations and remove the affect of the number of BS antennas. Compared to those iterative methods in [5] [6], the huge burden introduced by the number of BS antennas is greatly reduced by utilizing two small-size SVDs, which could greatly reduce the computation complexity.

➢ Thirdly, our methods cope with the demands of different number of user's streams. These parameters in these methods could be freely changed according to the realistic setting, with the constraints, such as the number of effective channel rank and BS/user's antennas, capacity performance and complexity.

The 3D channel model under Urban Micro cell with high (outdoor/indoor) UE density (3D-UMi) scenario calibrated by 3GPP [9] is adopted. In MU-MIMO system, the link-level performance of the proposed SVD-ZF methods and the complexity are analyzed and compared. The rest of this paper is organized as follows. Section II describes the problem formulation. The proposed low-complexity channel reconstruction methods are presented in Section III. Meanwhile, it outlines complexity analysis. The simulation is performed in Section IV. And the last Section gives the conclusion.

Notation: Bold letters are utilized to denote the matrix or the vector. $\mathbf{I}_K$ is the identity matrix of $K \times K$. $CN(\boldsymbol{m}, \mathbf{R})$ denotes the circular symmetric complex Gaussian distribution with mean $\boldsymbol{m}$ and covariance matrix $\mathbf{R}$. The superscripts T, H denote the transpose and conjugate transpose, respectively. The Kronecker product is denoted by $\otimes$.

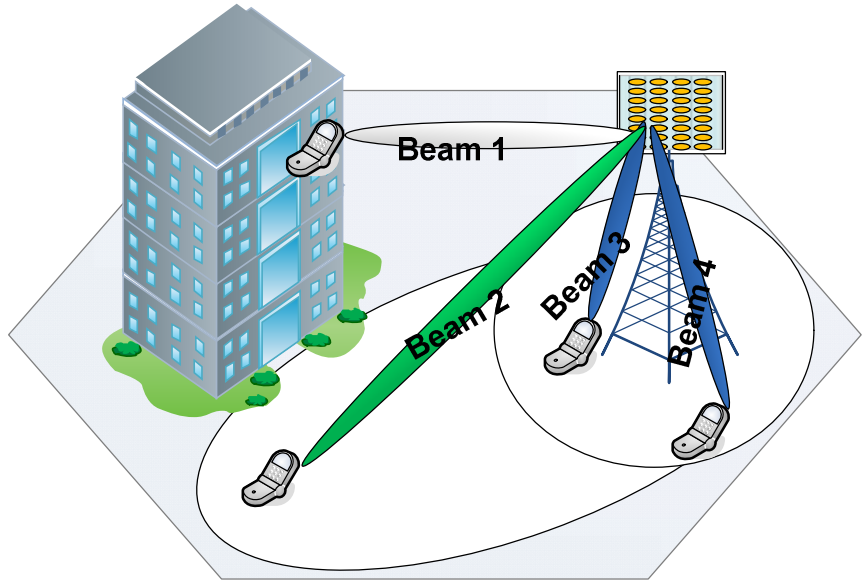

**Fig.1. Illustration of Massive 3D-MIMO systems: The elevation beams are divided into 3 parts; Beam 1 serves the users in high building; Beam 2 is pointing to users in far-subsector. The other beams belong to near-subsector users.**

## II. SYSTEM MODEL AND PROBLEM FORMULATION

### 2.1 3D channel model

Relative to the traditional 2D spatial channel model (SCM) channel model, the 3D channel model considers the radio propagation in the vertical dimension. Here the 3D channel model calibrated by the 3GPP is recommended [9].

The generation of the large scale parameters can be referred to [9]. The small-scale channel coefficient is generated by summing the contribution of some rays.

Assume that one ray is composed of $M_L$ sub-paths. The coordinate system for 3D channel model can be seen from Fig.2. $ray_{m,n}$ means the m-th sub-path in the n-th ray. The global coordinate system defines the zenith angle $\theta_{ZoD/ZoA}$ and the azimuth angle $\phi_{AoD/AoA}$. $\theta_{ZoD/ZoA} = 90^0$ points to the horizontal direction and $\theta_{ZoD} = 0°$ points to the zenith direction[9]. $\hat{n}$ is the given direction of $ray_{m,n}$. $\hat{\varphi}$ and $\hat{\theta}$ are the spherical basis vectors.

The channel coefficient from transmitter element s to receiver element u for the n-th ray is modeled as:

For NLOS path,

$$H_{u,s,n}(t) = \sqrt{P_n/M} \sum_{m=1}^{M} \begin{bmatrix} F_{rx,u,\theta}(\theta_{n,m,ZOA}, \varphi_{n,m,AOA}) \\ F_{rx,u,\varphi}(\theta_{n,m,ZOA}, \varphi_{n,m,AOA}) \end{bmatrix}^T \begin{bmatrix} \exp(j\Phi_{n,m}^{\theta\theta}) & \sqrt{\kappa_{n,m}^{-1}} \exp(j\Phi_{n,m}^{\theta\varphi}) \\ \sqrt{\kappa_{n,m}^{-1}} \exp(j\Phi_{n,m}^{\varphi\theta}) & \exp(j\Phi_{n,m}^{\varphi\varphi}) \end{bmatrix}$$

$$\begin{bmatrix} F_{tx,s,\theta}(\theta_{n,m,ZOD}, \varphi_{n,m,AOD}) \\ F_{tx,s,\varphi}(\theta_{n,m,ZOD}, \varphi_{n,m,AOD}) \end{bmatrix} \exp(j2\pi\lambda_0^{-1}(\hat{r}_{rx,n,m}^T \bar{d}_{rx,u})) \exp(j2\pi\lambda_0^{-1}(\hat{r}_{tx,n,m}^T \bar{d}_{tx,s})) \exp(j2\pi v_{n,m} t)$$

For LOS path,

$$H_{u,s,n}(t) = \sqrt{\frac{1}{K_R+1}} H_{u,s,n}^{'}(t)$$

$$+ \delta(n-1) \sqrt{\frac{K_R}{K_R+1}} \begin{bmatrix} F_{rx,u,\theta}(\theta_{LOS,ZOA}, \varphi_{LOS,AOA}) \\ F_{rx,u,\varphi}(\theta_{LOS,ZOA}, \varphi_{LOS,AOA}) \end{bmatrix}^T \begin{bmatrix} \exp(j\Phi_{LOS}) & 0 \\ 0 & -\exp(j\Phi_{LOS}) \end{bmatrix}$$

$$\begin{bmatrix} F_{tx,s,\theta}(\theta_{LOS,ZOD}, \varphi_{LOS,AOD}) \\ F_{tx,s,\varphi}(\theta_{LOS,ZOD}, \varphi_{LOS,AOD}) \end{bmatrix} \cdot \exp(j2\pi\lambda_0^{-1}(\hat{r}_{rx,LOS}^T \bar{d}_{rx,u})) \cdot \exp(j2\pi\lambda_0^{-1}(\hat{r}_{tx,LOS}^T \bar{d}_{tx,s})) \cdot \exp(j2\pi v_{LOS} t)$$

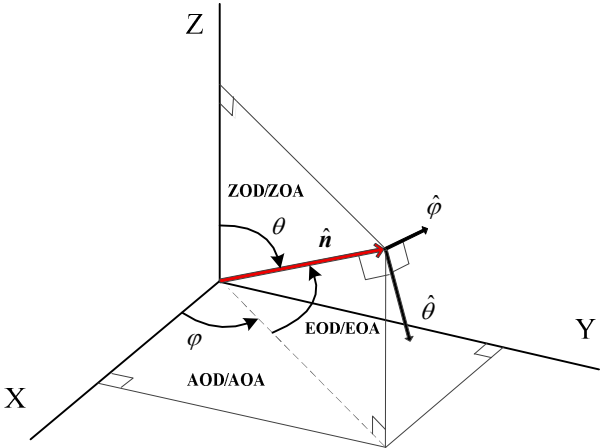

**Fig.2. The coordinate system for 3D channel model**

where $F_{rx,u,\theta}$ and $F_{rx,u,\varphi}$ are the antenna radiation patterns for element u in the direction of the spherical basis vectors, $\hat{\theta}$ and $\hat{\varphi}$ respectively. $F_{tx,s,\theta}$ and $F_{tx,s,\varphi}$ are the antenna radiation patterns for element s in the direction of the spherical basis vectors, $\hat{\theta}$ and $\hat{\varphi}$ respectively.

$$\hat{r}_{rx,n,m} = \begin{bmatrix} \sin\theta_{n,m,ZOA} \cos\varphi_{n,m,AOA} \\ \sin\theta_{n,m,ZOA} \sin\varphi_{n,m,AOA} \\ \cos\theta_{n,m,ZOA} \end{bmatrix}$$

is the unit vector about the azimuth of arrival angle(AoA)$\phi_{n,m,AOA}$ and the zenith of arrival angle(ZoA)$\theta_{n,m,ZOA}$. And also $\hat{r}_{tx,n,m}$ is the counterpart at the transmit side; $\bar{d}_{rx,u}$ and $\bar{d}_{tx,s}$ are the location vectors of the transmit and receive elements, respectively; $\{\Phi_{n,m}^{\theta\theta}, \Phi_{n,m}^{\theta\phi}, \Phi_{n,m}^{\phi\theta}, \Phi_{n,m}^{\phi\phi}\}$ are the random initial phases for sub-path m of ray n; $\lambda_0$ is the wavelength of the carrier frequency; $K_R$ is the Ricean K-factor; $v_{n,m}$ is the Doppler frequency component. More detailed description about the generation of 3D channel model can be referred to [9].

## 2.2 3D-MIMO system model

The BS in each cell is equipped with 2D planar cross-polarized antenna array. The number of antenna elements in azimuth direction and elevation direction is $N_A$ and $N_E$ respectively. And the total number of antennas is $Nt = N_A \times N_E \times 2$ with dual-polarized style. The configuration of antenna array is presented in Fig.3. Each user is equipped with $M$ antennas.

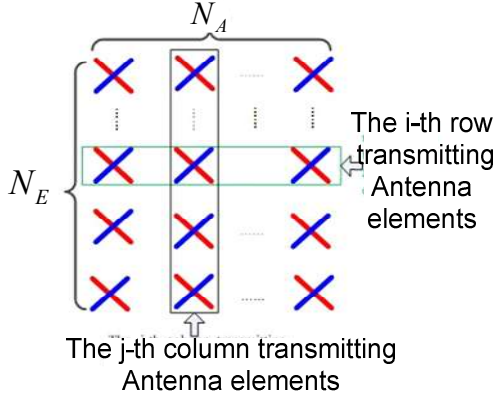

Fig.3. The coordinate system for 3D channel model

The received signal $\mathbf{x}^k \in C^{S_k \times 1}$ of the k-th user in MU-MIMO can be expressed as:

$$\mathbf{x}^k = \sqrt{\rho_f}\mathbf{E}^k_{(S_k \times M)}\mathbf{H}^k_{(M \times Nt)}\mathbf{W}^k_{(Nt \times S_k)}\mathbf{s}^k_{(S_k \times 1)} + \sum_{l=1,l \neq k}^{K}\sqrt{\rho_f}\mathbf{E}^k\mathbf{H}^k\mathbf{W}^l\mathbf{s}^l + \mathbf{v}_k$$

Where $M$, $Nt$, $S_k$ are the num of user antennas, the num of BS antennas and the allocated streams for k-th user, respectively. $\sum_{l=1,l \neq k}^{K}\sqrt{\rho_f}\mathbf{E}^k\mathbf{H}^k\mathbf{W}^l\mathbf{s}^l$ represents the interferences from the other users. $\mathbf{H}^k_{(M \times Nt)}$ is the k-th downlink channel matrix, K is the number of users, $\mathbf{W}^k_{(Nt \times S_k)}$ is the k-th precoding matrix. $\mathbf{E}^k_{(S_k \times M)}$ is the k-th estimating matrix. $\mathbf{s}^k_{(S_k \times 1)}$ is transmitted signal for the k-th user. $\mathbf{v}_k \sim CN(\mathbf{0}, \sigma^2\mathbf{I}_{S_k \times S_k})$ is the noise in the users.

**2.3 Problem formulation**

In conventional LTE systems, the users are usually equipped with $M$ antennas (e.g., 2, 4 or 8), and allocated $S$ layers (e.g., 1 or 2), which means that BS has to extract $S*Nt$ effective channels from $M*Nt$ estimated user channels before downlink precoding. In fact, the number of effective data steams is less than of user's antennas ($M \geq S$) in order to maintain the high performance. The whole process is showed in Fig.4. In TDD-MIMO systems, the users send sounding reference signals (SRS) to the BS in the uplink, and then, by these SRS, the BS estimates and gets the channels $\mathbf{H}_{M \times Nt}$ of different users. And how does the BS get the effective user channels from $M*Nt$ channel matrix to $S$ streams? The measure is to make channel reconstruction by the SVD in $H_{M \times Nt}$, which is to get the main singular-vector $H_{S \times Nt}$. Next, BS gets downlink effective channel $H_{S \times Nt}$ by TDD reciprocity and makes the downlink precoding. And the user takes the SVD to get the effective channels. Therefore, our focus challenge is the Step: 2 ($H_{M \times Nt} \to H_{S \times Nt}$).

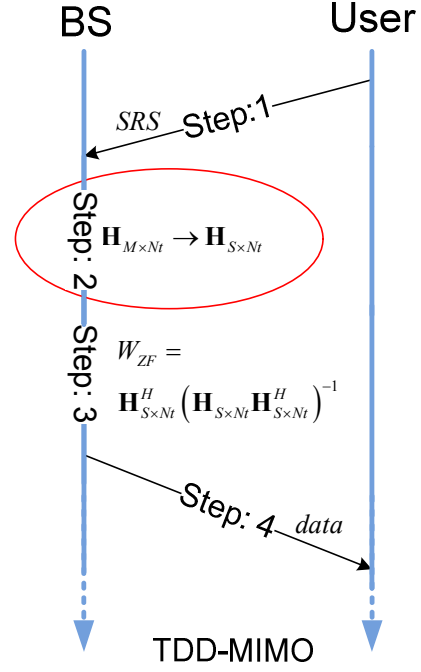

Fig.4. The illustration of the downlink precoding process. For TDD systems, step-1 completes users' SRS in uplink, and BS estimates uplink channel $H_{M \times Nt}$ with all users' antennas, and extracts effective channel $H_{S \times Nt}$ for downlink precoding in step-2. Next is to process precoding and to send data by $H_{S \times Nt}$.

**2.4 Traditional channel reconstruction method**

After getting the realistic user's MIMO channel by estimating the SRS, BS makes channel reconstruction with the transform from $H_{M \times Nt}$ to $H_{S \times Nt}$, based on per precoding unit (PU). One PU could be a resource block (RB), or many RBs, or across the whole bandwidth. If we define one PU includes $N_{RB}$ RBs, and one RB is with $N_{SC}$ subcarriers. So there are $N_{RB}N_{SC}$ subcarriers. Here, there is $V_A V_E$ antennas at BS, and $V_A$ and $V_E$ are the number of column and line respectively.

The traditional eigen-beamforming (EBF) method is to calculate the correlation matrix and to make the SVD, and here we call it *Direct SVD*.

| Direct SVD |
|---|
| S-1: Calculate the average correlation matrix **R** based on $N_{RB}N_{SC}$ channels per PU. $$\mathbf{R}^k_{(Nt \times Nt)} = \frac{1}{N_{RB}N_{SC}}\sum_{n}^{N_{RB}N_{SC}}(\mathbf{H}^{n,k}_{(M \times Nt)})^H(\mathbf{H}^{n,k}_{(M \times Nt)})$$ |
| S-2: Get the main singular-vectors and reconstruct the channel. $$[\mathbf{v}^{1,k}_{(Nt \times 1)}, \mathbf{v}^{2,k}_{(Nt \times 1)}, \ldots, \mathbf{v}^{S,k}_{(Nt \times 1)}] = svd\{\mathbf{R}^k_{(Nt \times Nt)}\}$$ $$(\hat{\mathbf{H}}_k)_{S \times Nt} = (\mathbf{v}^{1,k}_{(Nt \times 1)}, \mathbf{v}^{2,k}_{(Nt \times 1)}, \ldots, \mathbf{v}^{S,k}_{(Nt \times 1)})^H$$ |

In S-2, the core part is the SVD manipulation, which would produce more computation with $Nt$ increasing. So one potential solution is to reduce the dimension of the correlation matrix $\mathbf{R}$.

## III. PROPOSED METHODS

### 3.1 The source inspiration

With the increasing of BS antennas, the uniform planar arrays (UPAs) would replace traditional uniform linear arrays (ULAs) in future 5G. In UPAs, the two-dimensional antennas could be divided into elevation antennas and azimuth antennas, which hold different correlations. For example, there is much similar information between the adjacent columns or the adjacent lines of the UPAs in 3GPP 3D-MIMO model [9]. That is to say that we can extract main partial antenna information to approximate complete channel information. In math, the main singular-vector is considered. What's more, the spatial characteristics of antennas also affect the correlation. For instance, the same polarized columns of antennas share similar spatial information. Meanwhile, we note that the column includes more spatial information than the line antenna in UPAs, e.g., spatial extension of angle in elevation / azimuth and the number of spatial sub-paths, which means that line antennas take more spatial information and should be approximated more accurately.

So, we consider the optimization with the correlation of line / column antennas and different polarizations, and try to extract the main representative vectors. And utilize the limited information to replace the complete channel matrix.

### 3.2 Three different implementation methods

Here we can extract one column vector $\mathbf{v}_{(V_E \times 1)}$ and $S$ line vectors $\mathbf{v}_{(V_A \times 1)}$ to approach all of the column and line antennas information. And then, the Kronecker "$\otimes$" makes $\mathbf{v}_{(V_E \times 1)}$ and $\mathbf{v}_{(V_A \times 1)}$ into $(\hat{\mathbf{H}}_k)_{S \times Nt}$.

| Method I |
|---|
| S-1: Calculate the average elevation correlation matrix $\mathbf{R}$ based on $V_A$ column antennas vectors and $N_{RB}N_{SC}$ effective channels per PU. $$\mathbf{R}^k_{(V_E \times V_E)} = \frac{1}{N_{RB}N_{SC}V_A} \sum_{n}^{N_{RB}N_{SC}} \sum_{v}^{V_A} (\mathbf{H}^{n,k,v}_{(M \times V_E)})^H (\mathbf{H}^{n,k,v}_{(M \times V_E)})$$ where $\mathbf{H}^{n,k,v}_{(M \times V_E)}$ is the channel matrix of the k-th user at the v-th subcarrier per PU, corresponding to the n-th column of the BS array. |
| S-2: Calculate the main column singular-vector based on elevation antennas correlation matrix $\mathbf{R}^k_{(V_E \times V_E)}$ $$\mathbf{v}^{k,E}_{(V_E \times 1)} = svd\{\mathbf{R}^k_{(V_E \times V_E)}\}$$ |
| S-3: Calculate the equivalent azimuth channel $$\overline{\mathbf{H}}^{n,k}_{(M \times V_A)} = \mathbf{H}^{n,k}_{(M \times Nt)} \left[ \mathbf{I}_{V_A \times V_A} \otimes \mathbf{v}^{k,E}_{(V_E \times 1)} \right], \text{``} \otimes \text{''} \text{ is Kronecker multiplication}。$$ |
| S-4: Calculate the average azimuth correlation matrix $\mathbf{R}$ $$\mathbf{R}^k_{(V_A \times V_A)} = \frac{1}{N_{RB}N_{SC}} \sum_{n}^{N_{RB}N_{SC}} (\overline{\mathbf{H}}^{n,k}_{(M \times V_A)})^H (\overline{\mathbf{H}}^{n,k}_{(M \times V_A)})$$ |
| S-5: Calculate the main column singular-vector based on azimuth antennas correlation matrix $\mathbf{R}^k_{(V_E \times V_E)}$ $$[\mathbf{v}^{1,A,k}_{(V_A \times 1)}, \mathbf{v}^{2,A,k}_{(V_A \times 1)}, \ldots, \mathbf{v}^{S,A,k}_{(V_A \times 1)}] = svd\{\mathbf{R}^k_{(V_A \times V_A)}\}$$ |
| S-6: Reconstruct the channel by "$\otimes$" $\mathbf{v}^{i,A,k}_{(V_A \times 1)}$ and $\mathbf{v}^{k,E}_{(V_E \times 1)}$. $(\hat{\mathbf{H}}_k)_{S \times Nt} =$ $[\mathbf{v}^{1,A,k}_{(V_A \times 1)} \otimes \mathbf{v}^{k,E}_{(V_E \times 1)}, \mathbf{v}^{2,A,k}_{(V_A \times 1)} \otimes \mathbf{v}^{k,E}_{(V_E \times 1)}, \ldots, \mathbf{v}^{S,A,k}_{(V_A \times 1)} \otimes \mathbf{v}^{k,E}_{(V_E \times 1)}]^H$ |

Method I totally averages the elevation antennas information into one vector $\mathbf{v}^{k,E}_{(V_E \times 1)}$, which only requires one SVD with $V_E$. But there is much loss of elevation antennas information, which could cause bad spatial resolution in elevation. Next, Method II avoids averaging all the elevation antennas information, but introduces $V_A$ SVD computation.

| Method II |
|---|
| S-1: Calculate the average elevation correlation matrix $\mathbf{R}$ based on each column antenna vectors and $N_{RB}N_{SC}$ effective channels per PU. $$\mathbf{R}^{k,v}_{(V_E \times V_E)} = \frac{1}{N_{RB}N_{SC}V_A} \sum_{n}^{N_{RB}N_{SC}} (\mathbf{H}^{n,k,v}_{(M \times V_E)})^H (\mathbf{H}^{n,k,v}_{(M \times V_E)}), \quad v = 1, \ldots, V_A$$ |
| S-2: Calculate the main column singular-vector based on elevation antennas correlation matrix $\mathbf{R}^{k,v}_{(V_E \times V_E)}$ $$\mathbf{v}^{k,v,E}_{(V_E \times 1)} = svd\{\mathbf{R}^{k,v}_{(V_E \times V_E)}\}$$ |
| S-3: Calculate the equivalent azimuth channel $$\overline{\mathbf{H}}^{n,k}_{(M \times V_A)} = \mathbf{H}^{n,k}_{(M \times Nt)} \begin{bmatrix} \mathbf{v}^{k,1,E}_{(V_E \times 1)} & & & \\ & \mathbf{v}^{k,2,E}_{(V_E \times 1)} & & \\ & & \ddots & \\ & & & \mathbf{v}^{k,V_A,E}_{(V_E \times 1)} \end{bmatrix}$$ |
| S-4: The same with S-4 in Method I. |
| S-5: The same with S-5 in Method I. |
| S-6: Reconstruct the channel by "$\otimes$" $\mathbf{v}^{i,A,k}_{(V_A \times 1)}$ and $\mathbf{v}^{k,j,E}_{(V_E \times 1)}$. $(\hat{\mathbf{H}}_k)_{S \times Nt} =$ $\begin{bmatrix} \mathbf{v}^{1,A,k}_{(V_A \times 1)}(1) * \mathbf{v}^{k,1,E}_{(V_E \times 1)} & \mathbf{v}^{2,A,k}_{(V_A \times 1)}(1) * \mathbf{v}^{k,1,E}_{(V_E \times 1)} & \cdots & \mathbf{v}^{S,A,k}_{(V_A \times 1)}(1) * \mathbf{v}^{k,1,E}_{(V_E \times 1)} \\ \mathbf{v}^{1,A,k}_{(V_A \times 1)}(2) * \mathbf{v}^{k,2,E}_{(V_E \times 1)} & & & \vdots \\ \vdots & \vdots & & \vdots \\ \mathbf{v}^{1,A,k}_{(V_A \times 1)}(V_A) * \mathbf{v}^{k,V_A,E}_{(V_E \times 1)} & \cdots & \cdots & \mathbf{v}^{S,A,k}_{(V_A \times 1)}(V_A) * \mathbf{v}^{k,V_A,E}_{(V_E \times 1)} \end{bmatrix}^H$ where $\mathbf{v}^{1,A,k}_{(V_A \times 1)}(i)$ is the i-th element of $\mathbf{v}^{1,A,k}_{(V_A \times 1)}$. |

Method II nearly saves each column antennas information, but increases the complexity with additional SVD computation, compared to Method I. In the following, Method III focuses on different antenna polar styles.

*Method III*

S-1: Calculate the average elevation correlation matrix **R** based on column antenna vectors with the same polarization, and $N_{RB}N_{SC}$ effective channels per PU.

$$\mathbf{R}^{k,P_i}_{(V_E \times V_E)} = \frac{1}{N_{RB}N_{SC}V_A} \sum_{n}^{N_{RB}N_{SC}} \sum_{i \in S^{P_i}} (\mathbf{H}^{n,k,i}_{(M \times V_E)})^H (\mathbf{H}^{n,k,i}_{(M \times V_E)})$$

where $S^{P_i}$ is the set of column antenna vectors with same polar style $P_i$.

S-2: Calculate the main column singular-vector based on elevation antennas correlation matrix $\mathbf{R}^{k,P_i}_{(V_E \times V_E)}$

$$\mathbf{v}^{k,P_i,E}_{(V_E \times 1)} = svd\{\mathbf{R}^{k,P_i}_{(V_E \times V_E)}\}$$

S-3: Calculate the equivalent azimuth channel

$$\overline{\mathbf{H}}^{n,k}_{(M \times V_A)} =$$

$$\mathbf{H}^{n,k}_{(M \times Nt)} \begin{bmatrix} \mathbf{I}_{\frac{V_A}{N_P} \times \frac{V_A}{N_P}} \otimes \mathbf{v}^{k,P_1,E}_{(V_E \times 1)} & & \mathbf{0} \\ & \ddots & \\ \mathbf{0} & & \mathbf{I}_{\frac{V_A}{N_P} \times \frac{V_A}{N_P}} \otimes \mathbf{v}^{k,P_{N_P},E}_{(V_E \times 1)} \end{bmatrix}$$

where $N_P$ is the number of polarizations.

S-4: The same with S-4 in Method I.

S-5: The same with S-5 in Method I.

S-6: Reconstruct the channel by " $\otimes$ " $\mathbf{v}^{i,A,k}_{(V_A \times 1)}$ and $\mathbf{v}^{k,j,E}_{(V_E \times 1)}$.

$$(\hat{\mathbf{H}}_k)_{S \times Nt} = \begin{bmatrix} \mathbf{v}^{1,A,k}_{(V_A \times 1)}(1) * \mathbf{v}^{k,P_1,E}_{(V_E \times 1)} & \mathbf{v}^{2,A,k}_{(V_A \times 1)}(1) * \mathbf{v}^{k,P_1,E}_{(V_E \times 1)} & \cdots & \mathbf{v}^{S,A,k}_{(V_A \times 1)}(1) * \mathbf{v}^{k,P_1,E}_{(V_E \times 1)} \\ \mathbf{v}^{1,A,k}_{(V_A \times 1)}(2) * \mathbf{v}^{k,P_2,E}_{(V_E \times 1)} & & & \vdots \\ \vdots & & & \vdots \\ \mathbf{v}^{1,A,k}_{(V_A \times 1)}(V_A) * \mathbf{v}^{k,P_{N_{pA}},E}_{(V_E \times 1)} & \cdots & \cdots & \mathbf{v}^{S,A,k}_{(V_A \times 1)}(V_A) * \mathbf{v}^{k,P_{N_{pA}},E}_{(V_E \times 1)} \end{bmatrix}^H$$

Note that the sequence of the $\mathbf{v}^{k,j,E}_{(V_E \times 1)}$ in each column is depend on the realistic layout of different polarizations in the UPAs.

Method I II III shares the same idea that the one large dimension ($Nt$) SVD computation is divided into several small-size matrix decomposition. They require 2, $V_A+1$ and 3 SVD computation.

## 3.3 Complexity analysis

In this section, we analyze the computation complexity of the proposed methods and compare it with the complexity of the classical SVD (used in Direct SVD). We express the computation complexity in terms of the number of floating point operations (FLOPs). In the following, each scalar / complex addition or multiplication is counted as one FLOP. For the sake of simplicity, we do not distinguish between real-valued and complex-valued multiplications and neglect the computation complexity of common parts among these methods. Note that, although it cannot characterize the true computation complexity, FLOP counting captures the order of the computation load, so suffices for the purpose of the complexity analysis. Note that the exact number of calculation depends on the difference between implementation methods. Here, we take some popular calculation rules in [4] [10], which are widely used in methods analysis. The following table gives the necessary calculation conclusions of typical processes. Here, $\mathbf{A} \in \mathbb{C}^{M \times N}$, $\mathbf{C} \in \mathbb{C}^{M \times N}$, $\mathbf{B} \in \mathbb{C}^{N \times L}$ are arbitrary matrices. $\mathbf{U}$, $\mathbf{\Sigma}$, $\mathbf{V}$ are the corresponding decomposition matrix of $[\mathbf{U},\mathbf{\Sigma},\mathbf{V}] = svd(\mathbf{A})$. $\mathbf{Q}$ is the orthogonal matrix of $\mathbf{A}$ by QR decomposition.

**Table I** *Float computation of traditional matrix model*

| Matrix-Matrix Prod of $\mathbf{AA}^H$: | $M^2N + M(N - \frac{M}{2}) - \frac{M}{2}$ |
|---|---|
| Matrix-Matrix Prod of $\mathbf{AB}$ | $2MNL - ML$ |
| Matrix Scaling of $\alpha \mathbf{A}$ | $MN$ |
| Matrix-Matrix Sum of $\mathbf{A} + \mathbf{C}$ | $MN$ |
| QR Matrix decomposition for $\mathbf{A}$, required $\mathbf{Q}$ | $4(M^2N - MN^2 + \frac{N^3}{3})$ |
| SVD Matrix decomposition for $\mathbf{A}$, required $\mathbf{\Sigma}, \mathbf{U}$ | $4M^2N + 13N^3$ |
| SVD Matrix decomposition for $\mathbf{A}$, required $\mathbf{\Sigma}, \mathbf{V}$ | $2MN^2 + 13N^3$ |

For the proposed methods, the matrix decomposition holds makes up the most of the float computations. Meanwhile, the complexity order mainly depends on the dimension of the matrix, the times and the styles of decomposition method. The proposed Method I requires 2 SVDs respectively in elevation and azimuth dimensions. The Method II provides more accurate channel information for downlink precoding, while requiring $V_A+1$ SVDs. If considering dual-polarized antennas, we could utilize the proposed Method III, which requires 3 SVD computations.

Next we give an accurate float computation results, including matrix decomposition, addition and multiplication. Note that, the common part of float computation among these methods, and the manipulation beyond the methods' description, are ignored. By the conclusions in Table I, asymptotic complexity scaling is $O(Nt^2)$ for Direct SVD and $\max\{O(N_A^2), O(N_E^2)\}$ for Method I/II/III; thus, the asymptotic complexity gap between these methods is huge, especially for large $Nt$.

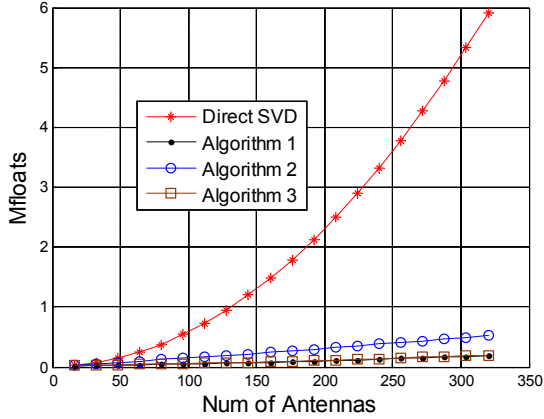

**Fig.5.** Complexity comparison between proposed Method I, II, III and Direct SVD, and the simulation parameters are based on the following Table II.

Fig.5 shows that, with the antennas increasing, larger dimension of matrix decomposition produces more float manipulations. Compared to Direct SVD, our proposed methods only require less than 10% float computations when the number of antennas is around 200 antennas, and a larger number of antennas could make the proposed methods more effective. Method II requires more float manipulations than other proposed methods. And the difference between Method II and III could be ignored. Considering the throughput in next section, the complexity is opposed to the performance, but there are some reasonable points, which make the balance between complexity and throughput.

## IV. SIMULATION RESULTS

In this section, we provide some numerical results to evaluate our proposed low-complexity channel reconstruction methods (Method I, Method II, Method III). The system model considered is in Section II with a BS employing a $8\times16$ UPA with 7 users, each having 8 antennas. The 3D-MIMO-UMi channels are modeled as [9]. Table II lists the detailed simulation parameters.

**Table II** *3D-MIMO simulation parameters*

| Parameters | Settings |
|---|---|
| Scenario | 3D-UMi |
| user antenna configuration | 8Rx, dual-polarized (0/+90) |
| BS antenna configuration ($H \times V$) | Dual-polarized $8H \times 8V$ |
| Bandwidth | 20MHz (100 RBs) |
| Antenna element interval | 0.5 carrier wave length both in horizontal and vertical direction |
| Carrier frequency | 2GHz |
| Number of users | 7 |
| Number of streams per user | 2 |
| MCS | Fixed, 64QAM |
| User distribution | Referred to [11] |
| User speed | 3km/h |
| Traffic model | Full buffer |
| Channel estimation | Ideal |
| Receiver | MMSE-IRC |

Considering the link simulation with MU-MIMO systems, Fig.6 exhibits the performance of the proposed methods with different normalization methods, which are based on entire precoding, per-user precoding and per-stream precoding respectively. Method II displays the best performance compared to other proposed methods. Besides, although Method I gets the worst performance, it is also around 1Gbps.

Whatever the normalization method is, Method II gets loss no more 10% than the ideal method (Direct SVD), meanwhile keeping low complexity. On the other hand, these methods make performance inversely to the computation complexity, just as showed in Fig.5 and Fig.6. So the final realization method should depend on the trade-off between the performance and complexity. Different normalization methods result in different performance, while all the throughputs are beyond 1Gbps. The normalization by per stream precoding provides more accurate interference information between different streams in zero-forcing precoding. Compared to the other normalization methods, it gets best performance.

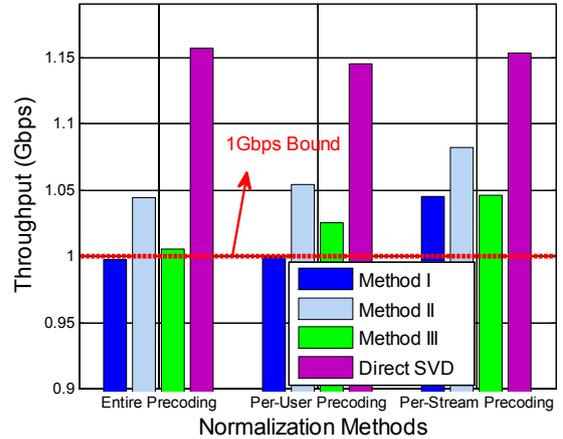

**Fig.6.** Achievable throughput using the proposed 3 methods while comparing to the ideal Direct SVD, with different normalization methods in 1 RB PU granularity.

If we consider correlation between adjacent resource block (RB), reasonable computational resolution could reduce the corresponding complexity. Here, we consider the different RB resolution cases by setting the parameter PU granularity $N_{RB}$, e.g. 1 RB, 2 RB and 4 RB in Fig.7. Similarly, we consider different PU granularities. Just as the analysis in Section III, PU granularity greatly affects the complexity computation. With 2 RB or 4 RB PU granularity, except Method II, both Method I and Method III are below 1Gbps. And the loss with Direct SVD is enlarged with PU granularity increasing.

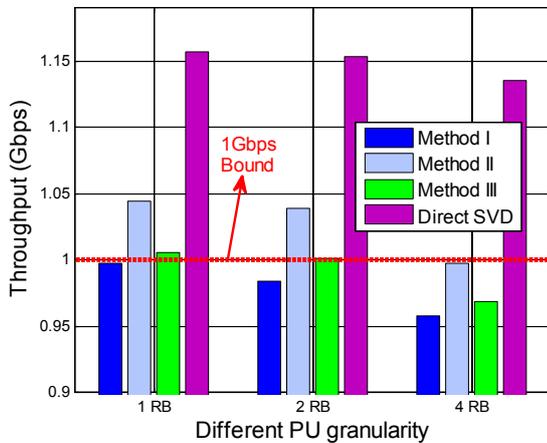

Fig.7. Achievable throughput using the proposed 3 methods and the ideal Direct SVD, with different PU granularity $N_{RB} = \{1, 2, 4\}$ **in entire precoding normalization.**

Based on the analysis and simulation results, we can conclude that the complexity and throughput are conflicted with each other. The larger PU granularity leads to less float computation, but the throughput gets less. In future 5G, while meeting the performance requirements, a reasonable complexity is necessary. From the above two figures, the affect of the normalization methods and the PU granularities is evaluated, and their joint optimization is necessary to cope with the demands of the performance and the computation complexity. The simulation results match our traditional expectation: higher resolution RB computation takes better performance; the higher complexity method gets better performance. And the simulation is based on realistic parameters in 3D-MIMO, and the results could be considered as a valuable performance target.

## V. CONCLUSIONS

This paper proposes a series of low-complexity channel reconstruction methods for TDD massive 3D-MIMO systems. These methods fully explore the correlation or the dependence between elevation and azimuth antennas, and divide the conventional large dimensional SVD into two stages of small-size SVD computation. Compared to the existing method, they could largely reduce the computation complexity, especially with large-scale antennas. The complexity analysis shows that the proposed methods only require less than 10% float computations, while the performance could be kept around 1Gbps by downlink ZF precoding.

## ACKNOWLEDGEMENT

This work was supported by the National High Technology Research and Development Program of China (863 Program) (Grant No. 2014AA01A705) and National Science and Technology Major Project of China (Grant No. 2015ZX03001034).

## Biographies

**Yuwei REN,** received the B.S. degree in communication engineering in 2011 and is currently pursing Ph.D. degree in communication and information systems from Beijing University of Posts and Telecommunications. His research interests include Massive MIMO, mmWave and Ad hoc networks. Especially he pays much attention into the design of practical implementation methods.

**Yang SONG,** received his Ph.D. degree from Beijing University of Posts and Telecommunications in 2006. He has worked with Alcatel-Lucent Shanghai Bell since 2006 and DOCOMO Beijing Communications Laboratories since 2010. Now he is with the State Key Laboratory of Wireless Mobile Communications, CATT, Beijing, China. His research interests are Massive MIMO system and its standardization in LTE/LTE-A.

**Xin SU,** received the B.S., M.S. and Ph.D. degrees from Xidian University, Xi-an, China in 2000, 2003 and 2006 respectively. From 2006 to 2007, he was with the System & Standard Department, Datang Mobile, CATT, Beijing, China. In 2007, he joined the System Lab, Telecommunication System Division, Samsung Electronics, Suwon, Korea. Since 2011, he has been with the State Key Laboratory of Wireless Mobile Communications, CATT, Beijing, China. His research interests are multi-antenna system and its standardization in LTE/LTE-A & IMT-2020(5G).